\documentstyle{mn}

\newif\ifAMStwofonts
\ifoldfss
  \ifCUPmtlplainloaded \else
    \NewTextAlphabet{textbfit} {cmbxti10} {}
    \NewTextAlphabet{textbfss} {cmssbx10} {}
    \NewMathAlphabet{mathbfit} {cmbxti10} {} % for math mode
    \NewMathAlphabet{mathbfss} {cmssbx10} {} %  "   "    "
  \fi
  \ifAMStwofonts
    \ifCUPmtlplainloaded \else
      \NewSymbolFont{upmath} {eurm10}
      \NewSymbolFont{AMSa} {msam10}
      \NewMathSymbol{\upi}     {0}{upmath}{19}
      \NewMathSymbol{\umu}     {0}{upmath}{16}
      \NewMathSymbol{\upartial}{0}{upmath}{40}
      \NewMathSymbol{\leqslant}{3}{AMSa}{36}
      \NewMathSymbol{\geqslant}{3}{AMSa}{3E}

       \let\ge=\geqslant
    \fi
  \fi
\fi % End of OFSS

\ifnfssone
  \newmathalphabet{\mathit}
  \addtoversion{normal}{\mathit}{cmr}{m}{it}
  \addtoversion{bold}{\mathit}{cmr}{bx}{it}
  \newmathalphabet{\mathbfit} % math mode version of \textbfit{..}
  \addtoversion{normal}{\mathbfit}{cmr}{bx}{it}
  \addtoversion{bold}{\mathbfit}{cmr}{bx}{it}
  \newmathalphabet{\mathbfss} % math mode version of \textbfss{..}
  \addtoversion{normal}{\mathbfss}{cmss}{bx}{n}
  \addtoversion{bold}{\mathbfss}{cmss}{bx}{n}
  \ifAMStwofonts
    \ifCUPmtlplainloaded \else
      \UseAMStwoboldmath
      \makeatletter
      \new@mathgroup\upmath@group
      \define@mathgroup\mv@normal\upmath@group{eur}{m}{n}
      \define@mathgroup\mv@bold\upmath@group{eur}{b}{n}
      \edef\UPM{\hexnumber\upmath@group}
      \new@mathgroup\amsa@group
      \define@mathgroup\mv@normal\amsa@group{msa}{m}{n}
      \define@mathgroup\mv@bold\amsa@group{msa}{m}{n}
      \edef\AMSa{\hexnumber\amsa@group}
      \makeatother
      \mathchardef\upi="0\UPM19
      \mathchardef\umu="0\UPM16
      \mathchardef\upartial="0\UPM40
      \mathchardef\leqslant="3\AMSa36
      \mathchardef\geqslant="3\AMSa3E

       \let\ge=\geqslant
    \fi
  \fi
\fi % End of NFSS release 1

\ifnfsstwo
  \DeclareMathAlphabet{\mathbfit}{OT1}{cmr}{bx}{it}
  \SetMathAlphabet\mathbfit{bold}{OT1}{cmr}{bx}{it}
  \DeclareMathAlphabet{\mathbfss}{OT1}{cmss}{bx}{n}
  \SetMathAlphabet\mathbfss{bold}{OT1}{cmss}{bx}{n}
  \ifAMStwofonts
    \ifCUPmtlplainloaded \else
      \DeclareSymbolFont{UPM}{U}{eur}{m}{n}
      \SetSymbolFont{UPM}{bold}{U}{eur}{b}{n}
      \DeclareSymbolFont{AMSa}{U}{msa}{m}{n}
      \DeclareMathSymbol{\upi}{0}{UPM}{"19}
      \DeclareMathSymbol{\umu}{0}{UPM}{"16}
      \DeclareMathSymbol{\upartial}{0}{UPM}{"40}
      \DeclareMathSymbol{\leqslant}{3}{AMSa}{"36}
      \DeclareMathSymbol{\geqslant}{3}{AMSa}{"3E}

       \let\ge=\geqslant
    \fi
  \fi
\fi % End of NFSS release 2

\ifCUPmtlplainloaded \else
  \ifAMStwofonts \else % If no AMS fonts
    \def\upi{\pi}
    \def\umu{\mu}
    \def\upartial{\partial}
  \fi
\fi

\input{epsfig.sty}

\title{Short Timescale Correlations Between Line and Continuum Fluxes in Cygnus X-1}
\author[T. J. Maccarone \& P. S. Coppi]
       {Thomas J. Maccarone \\
        Scuola Internationale Superiore di Studi Avanzati, via Beirut, n. 2-4, Trieste, Italy, 34014 
	\newauthor
	Paolo S. Coppi\\
	Department of Astronomy, Yale University, P.O. Box 208101, New Haven CT USA 06520-8101}
	
\date{}

\pagerange{\pageref{firstpage}--\pageref{lastpage}}
\pubyear{}
\begin{document}

\maketitle

\label{firstpage}

\begin{abstract}      
We present the results of 16-s time scale spectral fits for Cygnus X-1
in the soft state and in the transition state, using a Comptonised
blackbody plus an iron line.  On these timescales, we find that that
the continuum source flux can vary by factors 2-3 and that the iron
line intensity appears to track these changes well, i.e., the inferred
equivalent width of the line remains constant to within the errors.
We also find no significant changes in the seed (blackbody) photon
temperature, while the properties of the Comptonising corona clearly
do vary, with the spectral hardness and flux generally being
correlated.  The corona therefore seems to be the overall driver for
the rapid timescale variability observed in the soft and transition
states.  These results are consistent with the Fourier resolved
spectroscopy results of Gilfanov et al. (2000) that indicate the iron
line shows rapid flux variations while the blackbody component does
not and suggest that the iron line flux in fact tracks continuum
changes down to very short timescales.  We extend this work by showing
that not only the variability amplitudes, but also the {\it phases} of
the iron line and continuum components are identical.  We note that
the short timescale variability properties of the soft and transition
states are actually not very different from those of the hard state,
suggesting that the corona is the main cause of rapid variability in
that state too, and hence that the mechanism responsible for the
corona is similar in all three states.
\end{abstract}


\section{Introduction} 

The  spectra of  accreting black  holes  in the  soft state  generally
consist of a strong quasi-thermal component, generally assumed to come
from a  cold accretion  disc (Shakura \&  Sunyaev, 1973), a  power law
tail thought to originate in a corona of hot electrons above the disc,
and some additional features (a  Compton reflection bump and some iron
emission  lines)  thought to  result  from  reprocessing  of the  hard
photons  from the  corona by  the disc  (Basko, Sunyaev,  \& Titarchuk
1974; George \& Fabian 1991).

Several studies of active galactic nuclei (AGNs) have shown that
changes in the profile and equivalent width of the iron line can be
related to continuum changes (Iwasawa et al. 1996; Yaqoob et al. 1996;
Nandra et al. 1999; Wang et al. 1999; Weaver, Gelbord, \& Yaqoob
2000), suggesting that the iron line is indeed due to reprocessing of
the continuum by ambient cold matter, e.g., the accretion disc.  The
results of the AGN studies, however, have been somewhat ambiguous and
are likely complicated by factors that are less likely to arise in
compact binary systems.  Reprocessing from a dusty torus (Antonucci \&
Miller 1985) and the possibility that the iron line does not come
exclusively from the disc are two such factors (Weaver \& Reynolds
1998).  Studies of binaries thus should prove useful in providing
confirmation that the iron lines from accreting black holes really do
represent reprocessing of coronal X-rays by the disc and in
disentangling the disc components' contributions from other
contributions.

Timing characteristics have been used to provide evidence for the
applicability of reflection models in the canonical Galactic black
hole candidate Cygnus X-1 through the technique of Fourier resolved
spectroscopy (Gilfanov, Churazov, \& Revnivtsev, 2000).  This
technique works by Fourier transforming the lightcurve of Cyg X-1 in
several energy bands, then re-assembling a spectral energy
distribution at each Fourier frequency to determine the spectrum of
the component of the lightcurve that varies on a particular timescale.
Their analysis found that the reflected component and the primary
emission tracked each other down to timescale less than about 30 ms.
It also showed that the the iron line flux is roughly constant for
frequencies $\ge$ 30 Hz, giving an estimate for the inner disc size.

 A significant drawback of this technique, however, is that it loses
 phase information.  The iron line features cannot be definitively
 shown to be due to reflection unless some correlation is found
 between the phase of the iron line and the phase of the incident hard
 photons.  This paper seeks to complement the Fourier resolved
 spectrum by working instead in the time domain.  While photon
 statistics do not permit us to probe down to the very shortest
 timescales, we nonetheless can obtain useful spectral information
 down to short timescales $\sim 10$ seconds, where significant
 continuum flux variability is present.  By direct spectral fitting,
 we demonstrate that the iron line equivalent width remains constant
 on short timescales in both the soft state and the transition state
 of Cygnus X-1, i.e., that the iron line tracks continuum changes. At
 the same time, under the assumption that the observed continuum is
 Comptonised radiation, we find no evidence for significant changes in
 the seed photon distribution.  This suggests that the mechanism
 responsible for powering the hot coronal electrons is responsible for
 most of the observed flux variations, and that the disc is relatively
 quiescent.  In section 2 below, we outline our data reduction
 procedure.  In section 3 we discuss the fitting procedure and the
 model used.  In section 4, we show fits both for the 16 second
 timescales we emphasize in this work, and for the full RXTE dwells on
 the source in order to assess the problems that may be induced by
 fitting on the shorter timescales with the lower signal to noise.  In
 section 5, we compare short timescale variability in the soft and
 transition states to that in the hard state and show that it is
 likely that coronal changes drive the variability in all three
 states.  In section 6, we summarize all our conclusions.

\section{Observations} 

Analysis is presented of eight RXTE (Bradt, Rothschild, \& Swank,
1993) observations taken while Cygnus X-1 was in its ``soft state'' in
June of 1996 (proposal 10512).  The soft state of Cygnus X-1 shows a
stronger power law spectral tail and more variability than typical
black hole binary soft states and may actually be an intermediate
state (see Nowak 1995 for a review of spectral state nomenclature).
The data are prepared using the standard RXTE screening criteria:
earth elevation greater than 10 degrees, offset less than 0.01
degrees, all five proportional counter units on, and appropriate time
separation both before and after South Atlantic Anomaly Passages.
Spectra are then extracted on a 16-second time scale to allow for
background subtraction and response matrices are computed all using
the standard FTOOLS 5.0.4 routines.  All layers and all 5 Proportional
Counter Units are included, since the source dominates the background
and the most possible source counts are desired.  Response matrices
are computed using PCARMF version 7.10.  The total duration of the
data analyzed is 9856 seconds (9546 seconds of live time).  The
16-second time scale is chosen because it is convenient for background
subtraction purposes and because counting statistics do not permit
studies on much shorter time scales.

Results are also shown from the transition state into and out of the
soft state (RXTE proposal 10412).  These observations were taken in
May and August of 1996, respectively.  The same selection criteria
specified above are used to determine the good time intervals.  The
total duration of the transition state data analyzed is 14096 seconds
(13354 seconds of live time).  Finally, the hard state data from
December of 1997 (RXTE proposal 30158) were also analyzed, but the low
count rates do not allow for strong constraints to be place on the
variations of the equivalent widths of the iron lines in this state.

\begin{table*}
\begin{tabular}{ccc}
ObsID & Start Time & Stop Time \cr 
10412-01-02-00& 22/05/96 17:58:10& 22/05/96 19:49:13\cr
10412-01-01-00& 23/05/96 14:29:04& 23/05/96 18:08:13\cr
10412-01-03-00& 30/05/96 07:57:58& 30/05/96 08:45:13\cr
10512-01-07-00& 16/06/96 00:07:30& 16/06/96 00:41:13\cr
10512-01-07-02& 16/06/96 04:55:31& 16/06/96 05:44:13\cr 
10512-01-08-01& 17/06/96 01:44:32& 17/06/96 02:24:13\cr
10512-01-08-02& 17/06/96 04:56:32& 17/06/96 05:44:13\cr
10512-01-08-00& 17/06/96 08:08:32& 17/06/96 09:08:13\cr
10512-01-09-02& 18/06/96 03:21:34& 18/06/96 04:04:13\cr
10512-01-09-00& 18/06/96 06:34:33& 18/06/96 07:26:13\cr
10512-01-09-01& 18/06/96 09:46:33& 18/06/96 10:46:13\cr
10412-01-05-00& 11/08/96 07:21:46& 11/08/96 08:25:13\cr
10412-01-06-00& 11/08/96 15:25:58& 11/08/96 15:52:13\cr
\end{tabular}
\caption{The  data analyzed  for this  work.  Dates  are  presented in
DD/MM/YY format.  The observations from proposal 10512 are the soft
state observations while those from proposal 10412 are the transition
state observations.  The hard state data analyzed are all the
pointings from RXTE proposal 30158.}
\end{table*}

\section{Analysis}

\subsection{Discussion of Systematic Errors}

Observations of iron lines are not straightforward with an instrument
like RXTE that has poor energy resolution.  Iron line variability has
been seen in previous RXTE observations (e.g. Lee et al 2000), but
care must be taken with the data analysis to avoid spurious results.
In order to justify our choice of energy channels and to present
evidence that the correlations to be discussed are real and not
instrumental artefacts, we present spectral fits to the two Crab
observations taken most nearly in time to the Cyg X-1 observations
presented here.  In both cases, the neutral hydrogen column is fixed
to the Galactic value, $3.8 \times 10^{21}$ cm$^{-2}$.  The power law
index of the Crab is fit to be 2.175.  We then fit the normalisation
both with all channels included and with just the channels from 3.5-10
keV included.  For the Crab spectrum just after our observations (July
23, 1996), the deviations are less than 1\% in the 3.5-10 keV range
and generally less than 0.5\% when only the channels from 3.5-10 keV
are used for fitting the normalisation.  The residuals are slightly
larger for the observation taken before Cyg X-1 had begun its
transition to the high state (May 3, 1996).  The normalisations of the
two fits were consistent within much less than 1\%.  Since the
accuracy of the response matrix is significantly worse below 3.5 keV,
we ignore these channels.  We plot the results of the July 23 fit in
Figure 5.1.  The $\chi^2/\nu$ values for the two fits are 5.8 and 2.9
for the May 3 and July 23 fits respectively (with 17 bins and hence 15
degrees of freedom for each).  While these are formally quite bad
fits, the statistical errors for even a few hundred seconds of
observation of a source as bright as the Crab are quite small, and the
systematic errors dominate.  To estimate the appropriate systematic
errors, we add systematic errors to all channels and re-fit the data
until a $\chi^2/\nu$ of 1.0 is achieved.  We find that a 0.35\% error
is required for the May 3 observation and an 0.3 \% error is required
for the July 23 observation.  While this estimate of the systematic
errors is not strictly correct (as the systematic errors in the
different channels could be correlated), it represents the best
estimate that can be made with the given data.

\subsection{Spectral fits}

Using XSPEC 10.0 (Arnaud 1996), the spectra are fit from 3.5 keV to 10
keV (including 17 channels) with a simple model consisting of a
comptonised blackbody component (compbb in XSPEC - Nishimura, Mitsuda,
\& Itoh, 1986), a Gaussian component to fit the iron line.  The
spectral resolution of RXTE is not sufficient to provide strong
constraints on the shape of the feature we call an iron line, so other
explanations for this flux remain possible.  Still, it is the most
physically likely explanation for this excess flux, so we refer to
this feature as an iron line throughout the remainder of the paper.
This model was multiplied by photoelectric absorption (wabs).  When
more complicated models were fit to the data (such as pexrav -
Magdziarz \& Zdziarski 1995 to replace the power law component), the
increased number of free parameters resulted in degenerate fits and
the individual parameters had substantially larger error bars.  By
ignoring the channels above 10 keV, we are able to exclude the range
of energies where reflection is likely to be an important contributor
to the broad continuum features in the spectrum.  While using
complicated models over a larger range of energies might, in
principle, allow us to constrain the distribution of the Comptonising
electrons in the corona and the reflection fraction (as in Frontera et
al. 2000), such models are too computationally intensive for
applications to our short timescale data, and the lever arm from hard
photons needed to constrain the parameters of such models suffers from
poor statistics on short timescales.  The iron line center is
constrained to lie between 6.2 and 7.0 keV (but always fits best to
energies between 6.4 and 6.5 keV), the neutral hydrogen column density
is fixed at $5\times 10^{21}$ cm $^{-2}$, according to ASCA results
(see e.g., Gierlinski et al. 1999).  All the other parameters were
allowed to float freely.  A systematic error of 0.3\% is included as
suggested by the Crab fits.  We also plot the results of a ratio of
the source data to the Crab data in Figure 5.1b, and of a 16 second
integration in Figure 5.1c.

In Table 2 we present the results of spectral fits to the full
observations shown here, with an iron line included.  Since there are
17 energy channels, one can fit the 7 parameter model with 10 degrees
of freedom.  We also tabulate the $\Delta\chi^2$ that results from
removing the iron line from the model.  Fits of a reflection model
(i.e. pexriv) with no Gaussian component are similarly bad.

A more phenomenological model (disc blackbody + power law + Gaussian)
gives an acceptable fit in terms of $\chi^2/\nu$, but results in an
artificially low line energy ($\sim$ 6.2 keV) and an artificially high
line equivalent width ($>$ 1 keV).  This problem is relatively severe
because the two continuum components intersect near 6 keV.  The real
continuum of a Comptonised spectrum shows strong curvature at energies
where the seed blackbody spectrum begins to fall sharply, and the
phenomenological model fails to reproduce that curvature.  The
Gaussian component drifts to lower energy and to higher flux level to
fill in the missing flux from the incorrect estimation of the
continuum.  If the flux at the lowest energies contains a significant
contribution from Comptonised radiation, as suggested by the more
detailed long-timescale fits of Gierlinski et al.  (1999) and Frontera
et al (2000), then the temperature of the soft component is also
systematically overestimated by phenomenological model, as the peak
energy in the spectrum is shifted upwards by the Comptonisation
process driving the best fit temperature to larger values as well.
Using a Comptonised blackbody as the continuum model is not
computationally expensive and avoids all these problems. As we shall
see, doing so also gives parameters in reasonable agreement with those
obtained using the more detailed, wider energy range fits.

\begin{table*}
\begin{tabular}{cccccccccccc}
ObsID & BB Temp & $KT_{cor}$ & $\tau$ & LE & LW & LN & $\chi^2/\nu$ & $\Delta\chi^2$ & EqW & Flux\cr 

10412-01-01-00 & 0.43${\pm_{.10}^{.12}}$ & 21.1 ${\pm_{1.8}^{5.5}}$ & 1.53 ${\pm_{.77}^{.16}}$ & 6.46 ${\pm_{.10}^{.05}}$ & 0.71${\pm_{.06}^{.14}}$ & 4.4${\pm_{0.4}^{1.8}} \times 10^{-2}$ & 0.30 & 2600 & 230 & 14\cr
10412-01-02-00 & 0.37${\pm_{.06}^{.11}}$ & 25.1 ${\pm_{0.9}^{0.8}}$ & 0.83 ${\pm_{.10}^{.06}}$ & 6.46 ${\pm_{.06}^{.03}}$ & 0.66${\pm_{.05}^{.07}}$& 4.2${\pm_{.3}^{.6}} \times 10^{-2}$ & 0.23 & 1700 & 242 &13\cr
10412-01-03-00 & 0.38${\pm_{.10}^{.12}}$ & 26.1 ${\pm_{1.8}^{3.0}}$ & 0.73 ${\pm_{.46}^{.14}}$ & 6.47 ${\pm_{.05}^{.06}}$ & 0.70${\pm_{.06}^{.07}}$& 4.5${\pm_{.4}^{.5}} \times 10^{-2}$ & 0.55 & 2300 & 269 & 9.0\cr
10412-01-05-00 & 0.42${\pm_{.10}^{.11}}$ & 21.8 ${\pm_{1.5}^{1.6}}$ & 1.24 ${\pm_{.06}^{.12}}$ & 6.43 ${\pm_{.07}^{.06}}$ & 0.73${\pm_{.06}^{.09}}$& 4.9${\pm_{.5}^{.8}} \times 10^{-2}$ & 0.46 & 2200 & 275 & 14\cr
10412-01-06-00 & 0.41${\pm_{.08}^{.11}}$ & 22.6 ${\pm_{1.3}^{1.4}}$ & 1.03 ${\pm_{.11}^{.13}}$ & 6.38 ${\pm_{.09}^{.07}}$ & 0.75${\pm_{.08}^{.09}}$& 5.9${\pm_{0.7}^{1.0}} \times 10^{-2}$ & 0.16 & 1900 & 261 &17\cr
10512-01-07-00 & 0.38${\pm_{.09}^{.08}}$ & 26.1 ${\pm_{1.1}^{2.5}}$ & 0.61 ${\pm_{.15}^{.10}}$ & 6.41 ${\pm_{.08}^{.04}}$ & 0.74${\pm_{.08}^{.09}}$& 5.0${\pm_{0.9}^{2.7}} \times 10^{-2}$ & 0.47 & 2050 & 361 &11\cr
10512-01-07-02 & 0.37${\pm_{.05}^{.07}}$ & 26.7 ${\pm_{0.9}^{0.6}}$ & 0.51 ${\pm_{.15}^{.07}}$ & 6.41 ${\pm_{.09}^{.05}}$ & 0.75${\pm_{.08}^{.17}}$& 3.8${\pm_{0.7}^{2.0}} \times 10^{-2}$ & 0.52 & 2150 & 365 &9.0\cr
10512-01-08-00 & 0.38${\pm_{.09}^{.08}}$ & 26.1 ${\pm_{1.1}^{2.5}}$ & 0.61 ${\pm_{.15}^{.10}}$ & 6.41 ${\pm_{.08}^{.04}}$ & 0.74${\pm_{.08}^{.09}}$& 5.0${\pm_{0.9}^{2.7}} \times 10^{-2}$ & 0.47 & 2700 & 344 &15\cr
10512-01-08-01 & 0.37${\pm_{.08}^{.09}}$ & 27.9 ${\pm_{1.0}^{3.5}}$ & 0.46 ${\pm_{.16}^{.10}}$ & 6.37 ${\pm_{.10}^{.08}}$ & 0.77${\pm_{.10}^{.08}}$& 2.9${\pm_{0.7}^{2.5}} \times 10^{-2}$ & 0.51 & 1200 & 369 &6.5\cr
10512-01-08-02 & 0.38${\pm_{.09}^{.08}}$ & 26.1 ${\pm_{1.1}^{2.5}}$ & 0.61 ${\pm_{.15}^{.10}}$ & 6.41 ${\pm_{.08}^{.04}}$ & 0.74${\pm_{.08}^{.09}}$& 5.0${\pm_{0.9}^{2.7}} \times 10^{-2}$ & 0.47 & 2550 & 398 & 10\cr
10512-01-09-00 & 0.35${\pm_{.06}^{.05}}$ & 28.0 ${\pm_{0.7}^{1.5}}$ & 0.37 ${\pm_{.11}^{.08}}$ & 6.49 ${\pm_{.12}^{.06}}$ & 0.73${\pm_{.07}^{.07}}$& 2.3${\pm_{0.3}^{0.6}} \times 10^{-2}$ & 0.35 & 2200 & 356 &5.8\cr
10512-01-09-01 & 0.38${\pm_{.09}^{.08}}$ & 26.1 ${\pm_{1.1}^{2.5}}$ & 0.61 ${\pm_{.15}^{.10}}$ & 6.41 ${\pm_{.08}^{.04}}$ & 0.74${\pm_{.08}^{.09}}$& 5.0${\pm_{0.9}^{2.7}} \times 10^{-2}$ & 0.27 & 3150 & 367 &13\cr
10512-01-09-02 & 0.38${\pm_{.09}^{.08}}$ & 26.1 ${\pm_{1.1}^{2.5}}$ & 0.61 ${\pm_{.15}^{.10}}$ & 6.41 ${\pm_{.08}^{.04}}$ & 0.74${\pm_{.08}^{.09}}$& 5.0${\pm_{0.9}^{2.7}} \times 10^{-2}$ & 0.87 & 1600 & 411 &5.7\cr
\cr

\end{tabular}

\caption{The best fit parameters (along with their 90\% error intervals) for the full integrated observations of Cygnus X-1.  The columns names are, in order, Observation ID number, comptonised blackbody temperature in keV, temperature of the corona in keV, the optical depth of the corona, line energy, line width, line normalization in XSPEC units, $\chi^2/\nu$ for the best fits, $\Delta\chi^2$ when the iron line is removed, the equivalent width of the best fit iron line in eV, and the integrated flux from 3.5 to 10 keV in  $10^{-9}$   ergs/cm$^2$/sec.}
\end{table*}

\begin{figure*}
\centerline{\epsfig{figure=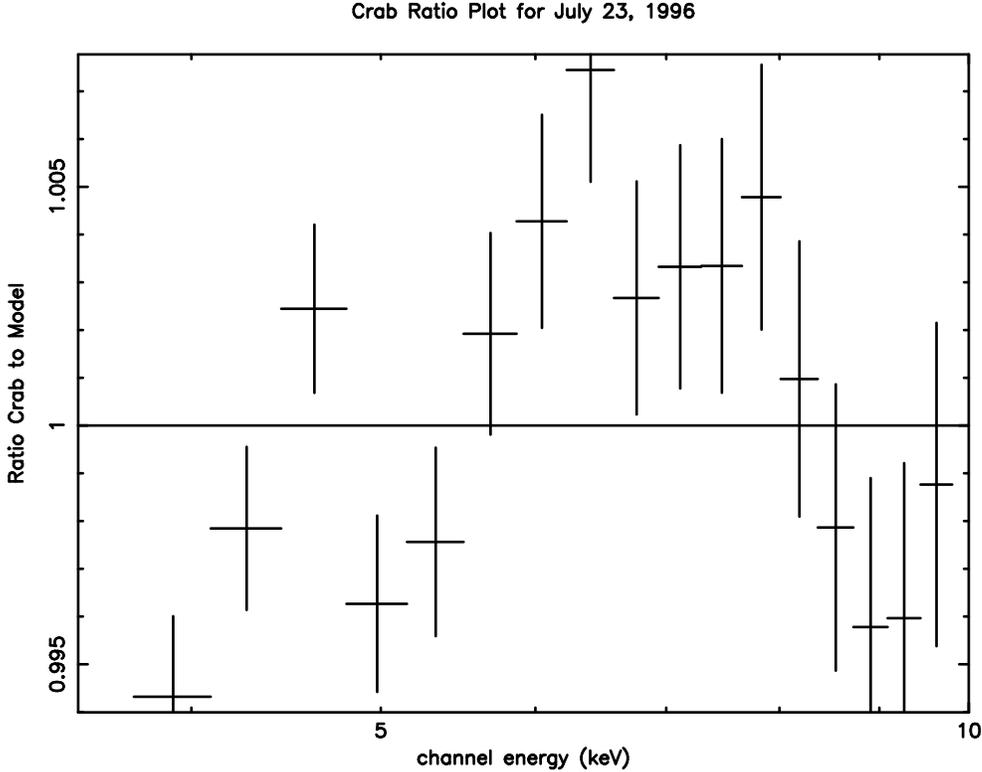,width=13cm,angle=-90}}
\caption{ 
The ratio of the observed Crab spectrum from July 23, 1996 to the best
fit with the power law index and neutral hydrogen column fixed.}
\end{figure*}

\begin{figure*}
\centerline{\epsfig{figure=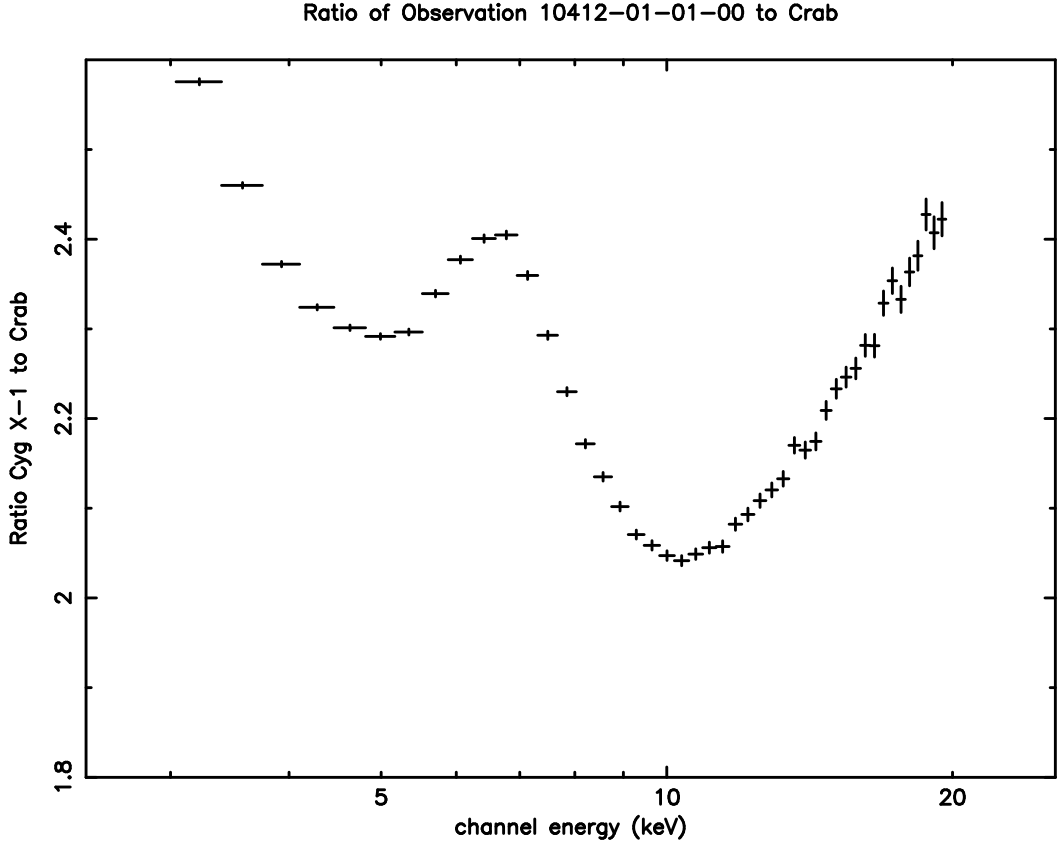,width=14cm,angle=-90}}
\caption{The ratio of observation 10412-01-01-00 to the same Crab
spectrum.  A strong feature is clearly seen peaking between 6 and 7
keV, so the iron line is unlikely to be a systematic feature.  The
plotted errors are 1$\sigma$ errors.}
\end{figure*}

\begin{figure*}
\centerline{\epsfig{figure=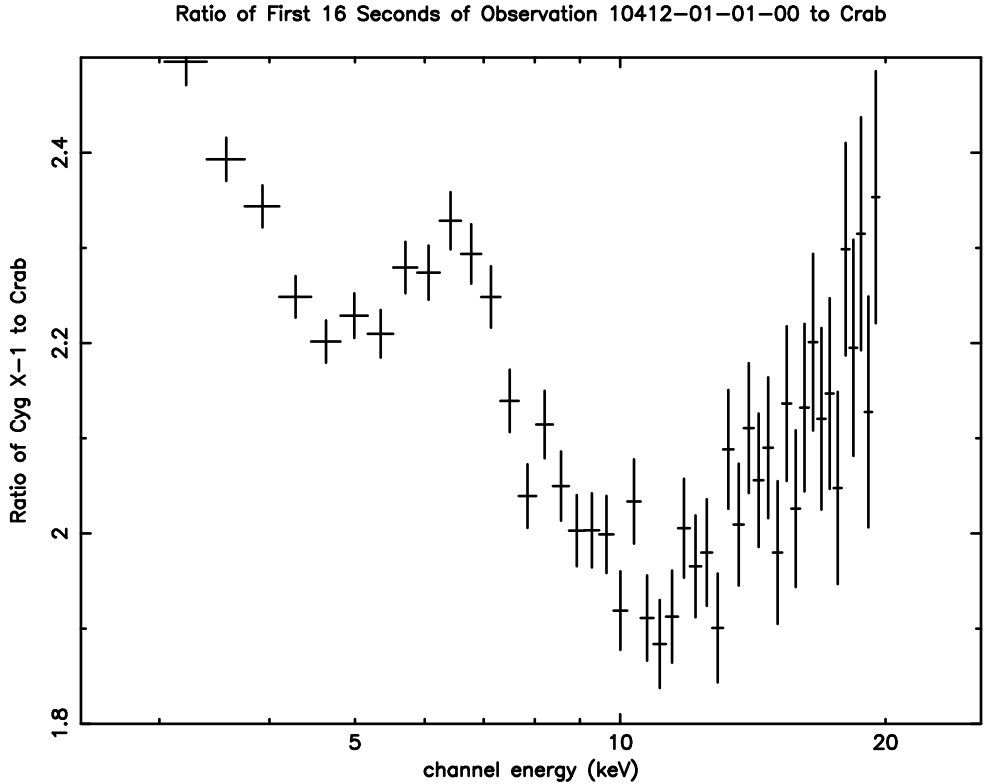,width=13cm,angle=-90}}
\caption{ The ratio of the first 16 second segment of observation
10412-01-01-00 to the same Crab spectrum.  Note that the shape is
essentially the same as in the previous figure.  The plotted errors
are 1 $\sigma$ errors.}
\end{figure*}

\begin{figure*}
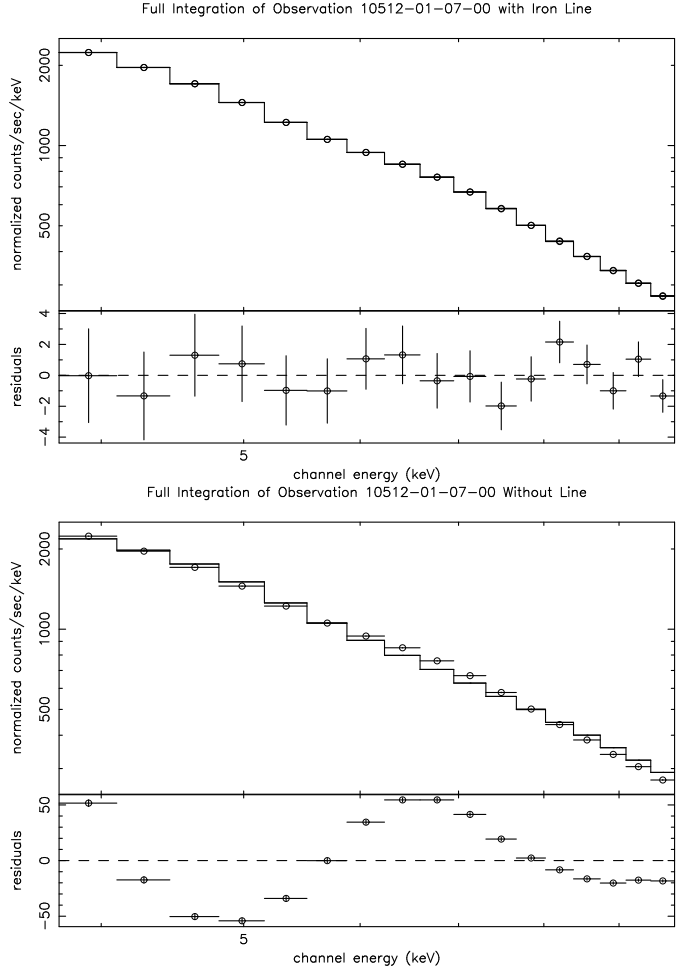

\centerline{\epsfig{figure=fullwithline.ps,width=3.5in,angle=-90}}
\centerline{\epsfig{figure=fullwoline.ps,width=3.5in,angle=-90}}
\caption{The full integration of Proposal 10512-01-07-00. Top: Fit and
residuals with line.  Bottom: Fit and residuals without line.  The
errors shown in the residuals represent only the statistical errors.
Small circles are used in both plots to identify the actual data
points, while the model is shown by the solid line.  Where the error
bars are not seen, they are smaller than the width of the line showing
the model.  The plotted errors are 1 $\sigma$ errors.}
\end{figure*}

\begin{figure*}
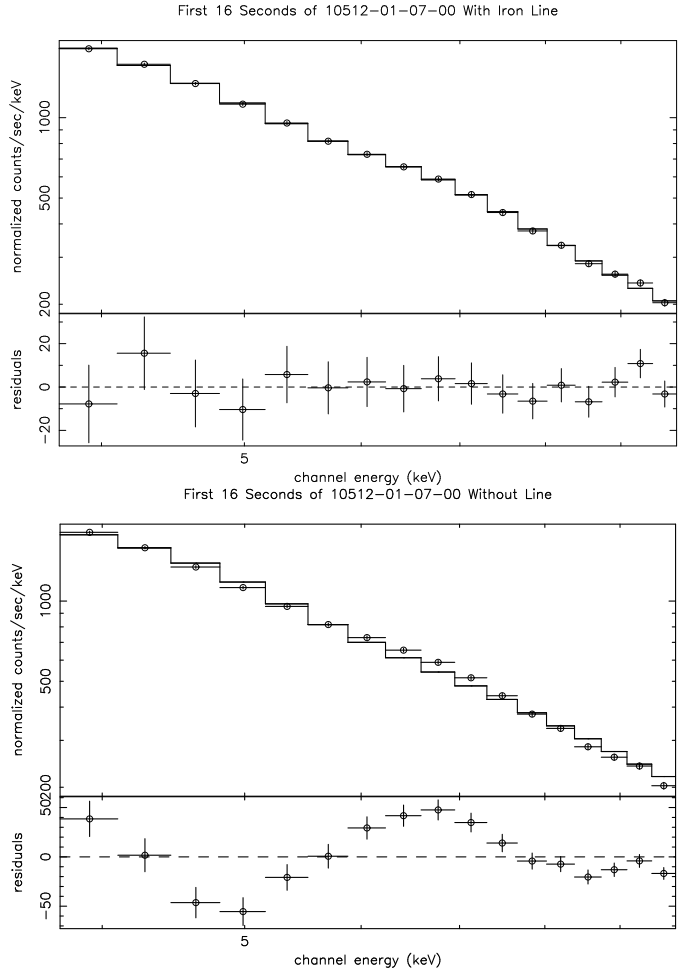

\centerline{\epsfig{figure=plot16withline.ps,width=3.5in,angle=-90}}
\centerline{\epsfig{figure=plot16woline.ps,width=3.5in,angle=-90}}
\caption{The fit to the first 16 seconds of Proposal
10512-01-07-00. Top: Fit and residuals with line.  Bottom: Fit and
residuals without line.  The $\Delta\chi^2$ for the two observations
is 329, so the iron line is extremely significant even on this short
timescale. The actual data points are indicated by small circles, and
the best fitting model is shown by the solid line.  Where the error
bars cannot be seen, they are smaller than the width of the line
showing the model.  The plotted errors are 1 $\sigma$ errors.}
\end{figure*}

\section{Results}

\subsection{Transition state - full integrations}

In the transition state, the blackbody temperature is always fit by a
value consistent with 400 eV. The value of the disc temperature as
well as the values of all the line parameters (line energy, physical
width and equivalent width) are consistent with those found from joint
fits to ASCA/RXTE observations taken during the same transition state
(Frontera et al., 2001) within the 90\% error bars, which lends
confidence to the interpretation that the continuum is well modeled by
compbb in the energy range examined and that the response of RXTE is
sufficient for making these measurements (since the ASCA observations
will place much stronger constraints on the parameters of line
features than RXTE does).  The iron line equivalent width seems to
rise going into the soft state, as the two lowest equivalent width
observations were taken during the earliest part of the transition
state, but the equivalent widths of the line in the different
transition state observations are all consistent with one another at
the 90\% confidence level, so no strong claims can be made.

There is evidence for variability of both the optical depth and the
electron temperature of the corona.  Since the best fit electron
temperatures are typical factors of $\sim$ 2-3 greater than the
highest photon energy used in the fit, the temperature and optical
depth are somewhat degenerate parameters, and small errors in either
the response matrix or the model itself could result in significant
deviations from the real values of the optical depth and temperature.
Furthermore, the soft state and the transition state of Cygnus X-1
have been shown to have strong non-thermal electron components
(Gierlinski et al.  1999; Frontera et al.  2001), so the physical
interpretation of the optical depth and temperature derived from the
best-fit thermal Comptonisation model is not clear.  Thus while the
observations suggest that the changes in the optical depth of the
system are more important than the changes in the temperature, one can
only be certain that the Compton $y$ parameter,
$$
	y = (4 k_B T)/(m_ec^2) Max(\tau,\tau^2)
$$
has changed, and not which component has actually changed.

\subsection{Soft state - full integrations}

In the soft state, the best fitting blackbody temperature is about 380
eV, slightly lower than the best fit for the transition state, but
consistent with having the same value as in the transition state.  The
iron lines show a slightly greater equivalent width than in the
transition state, typically around 370 eV, consistent with BeppoSax
observations of the soft state (Frontera et al.  2001).  The most
significant difference between the fit parameters to the integrated
spectra of the two states is that the optical depths are
systematically lower in the soft state than in the transition state.
As noted above, this can safely be interpreted only as a
lowering of the Compton $y$ parameter, i.e., a smaller typical
amplification of the energy of a seed photon in the soft state than in
the transition state.

\subsection{Short integrations}
\subsubsection{Procedure}

Having established an appropriate model for the long integrations, we
then fit this model to the data for the 16 second segments.  The
standard 0.3 \% systematic errors are added to each channel.  The iron
line energy, iron line physical width and the coronal temperature are
frozen to the values from the full integrations (leaving 13 degrees of
freedom), but even when allowed to vary, they typically differ by no
more than $\sim$ 10\% different than the values from the full
integrations.  These parameters are frozen because the fits are least
sensitive to them and because the $\chi^2$ values that are
significantly below 1 for the full integrations suggest the model is
``overfitting'' the data.  The optical depth parameter is allowed to
float rather than the coronal electron temperature simply because the
optical depth is observed to vary more dramatically in the best fits,
and not for any physically motivated reason.  The two parameters are
essentially degenerate given the energy range being fit.

\subsubsection{Results}
The $\chi^2/\nu$ values for the 16 second fits are typically about 1.0
(the mean $\chi^2/\nu$ is 1.01 for the soft state and 0.94 for the
transition state), and never larger than 2.8, nor smaller than 0.22.
Since over 1000 fits have been done, a few outliers at relatively
large and relatively small $\chi^2/\nu$ are to be expected.  Errors in
the best fit parameters are estimated using the steppar command within
XSPEC on a representative sample of the fits (direct computations of
all the errors is not computationally feasible for this large number
of fits).  Typical 90\% errors in the iron line flux level are about
20\%, errors in the optical depth are 5\%, and errors in the blackbody
temperature are 25\%.

We then look for the amplitude of variations of the different fit
parameters and for correlations in their variations.  In table 3, the
fractional RMS variations of the fit parameters are presented.  Since
the 90\% errors correspond to 1.65 $\sigma$ variations, the random
errors expected are about 12\% in iron line flux, 3\% in optical
depth, and 15\% in blackbody temperature.  One cannot explicitly fit
the equivalent width in XSPEC, so one cannot derive explicit error
bars, but they should be slightly larger than those of the iron line
flux normalization, since there are errors in both the flux of the
line and of the underlying continuum.  The errors in the overall flux
should be quite small ($<$ 1\%).

It is thus clear that the line normalisation, the overall flux, and
$\tau$ vary on 16 second timescales.  The line normalisation tracks
the overall flux, while the equivalent width does not.  In all the
observations except 10412-01-01-00, the iron line normalisation tracks
very well with the best fit value of $\tau$.  A typical plot (from
observation 10512-01-09-01) is shown in Figure 5.4a.  Since the
optical depth is a proxy for the spectral index of the power law, and
the line normalization correlates with the continuum flux, what this
correlation really shows is that in general, the spectrum gets harder
as the source gets brighter.  Why observation 10412-01-01-00 does not
fit the trend is unclear.  The errors for the best fit $\tau$ and $k_B
T_{cor}$ to the full integration are much larger than for the other
observations, so the most likely explanation is that the initial
parameters for the short timescale fits are somehow in error, but a
real physical difference between this observation and the others
cannot be ruled out.  However, the optical depth does not correlate
with the equivalent width, as is shown in Figure 5.4b.  The maximum
variations in the flux are factors of 2.6 within individual
observations (in obs 10512-01-08-00), 6 across the whole set of soft
state observations, and 3 across the whole set of transition state
observations.

The blackbody temperature and the iron line equivalent width show
variability that is generally consistent with or less than the fitting
errors.  In the cases where the variability is less than that expected
due to the errors, the likely cause is that the systematic errors
contribute to the size of the fitting errors, but due not contribute
to the variability.  We thus confirm previous results (Gilfanov,
Churazov \& Revnitsev 2000) that the seed photon component shows
little if any intrinsic variability on relatively short timescales.
Since these two components show no evidence of variability, any
correlations between their values and those of other parameters would
likely be due to errors in the fit parameters rather than intrinsic
correlations, so we do not discuss any such possible correlations.  We
do plot the iron line equivalent width versus flux (3.5 to 10 keV) for
observation 10412-01-01-00 in Figure 5.5.  These results are typical
of all the observations and demonstrate that there is no correlation
between line equivalent width and overall flux.  Furthermore, one can
see from the plot that the variations in the equivalent width are
consistent with being due to the fit errors.

\begin{table*}
\begin{tabular}{cccccc}
ObsID & $T_{BB}$& $\tau$& $LN$ & EqW & Flux \cr 
10412-01-01-00& .158&.309&.216&.124&.158\cr
10412-01-02-00& .086&.058&.139&.126&.084\cr
10412-01-03-00& .105&.073&.153&.145&.127\cr
10412-01-05-00& .220&.204&.152&.110&.081\cr
10412-01-06-00& .161&.118&.189&.173&.133\cr
10512-01-07-00& .014&.063&.229&.110&.189\cr
10512-01-07-02& .015&.071&.211&.175&.181\cr
10512-01-08-01& .103&.071&.242&.120&.222\cr
10512-01-08-02& .095&.122&.238&.163&.169\cr
10512-01-08-00& .100&.071&.195&.141&.149\cr
10512-01-09-00& .093&.135&.215&.148&.169\cr
10512-01-09-01& .093&.059&.219&.128&.191\cr
10512-01-09-02& .100&.133&.187&.167&.101\cr

\end{tabular}
\caption{The fractional variances of each fit parameter for the 16 second segment within the observations.}
\end{table*}

\begin{figure*}
\centerline{\epsfig{figure=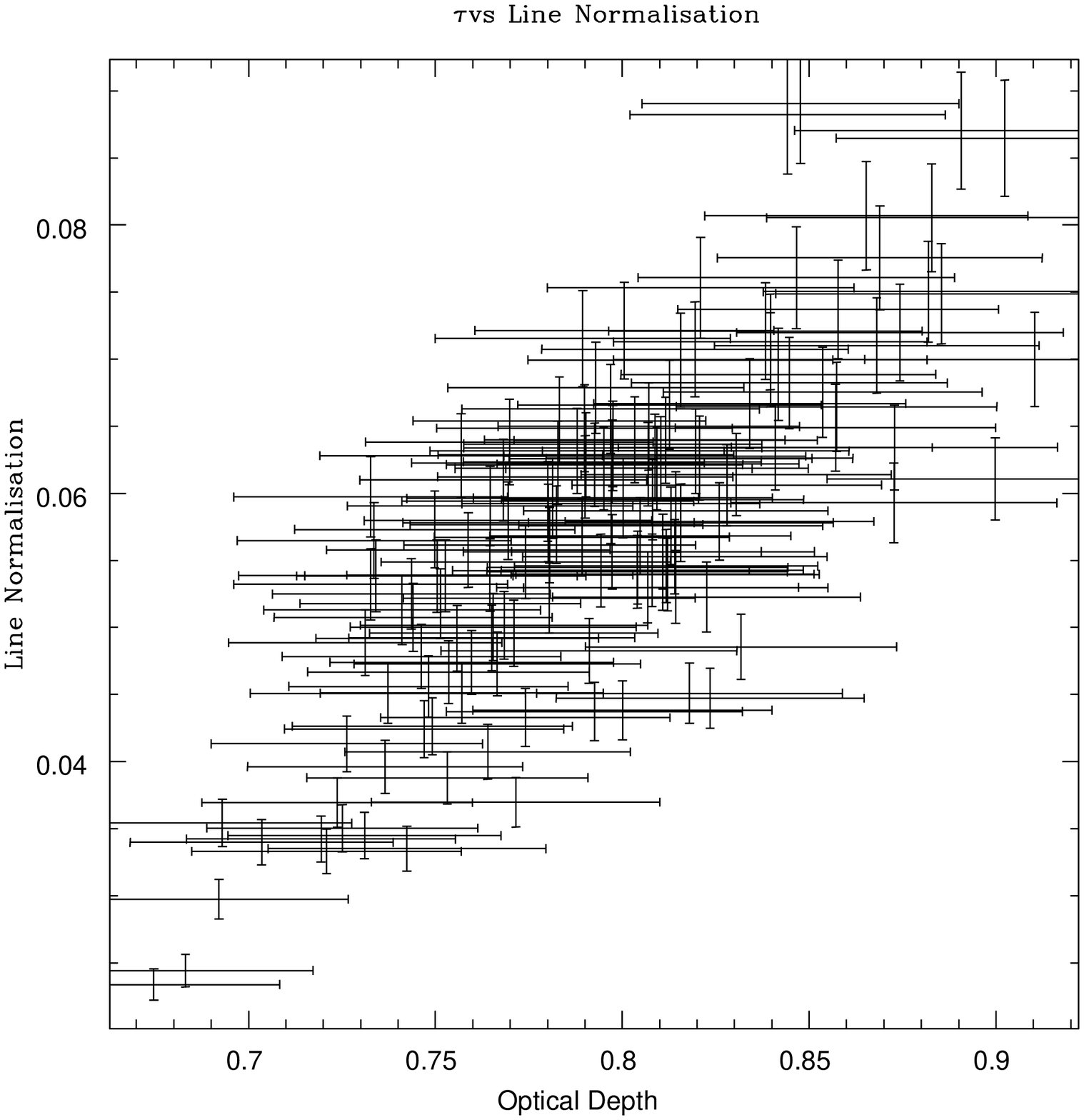,width=3.5in,angle=0}}
\centerline{\epsfig{figure=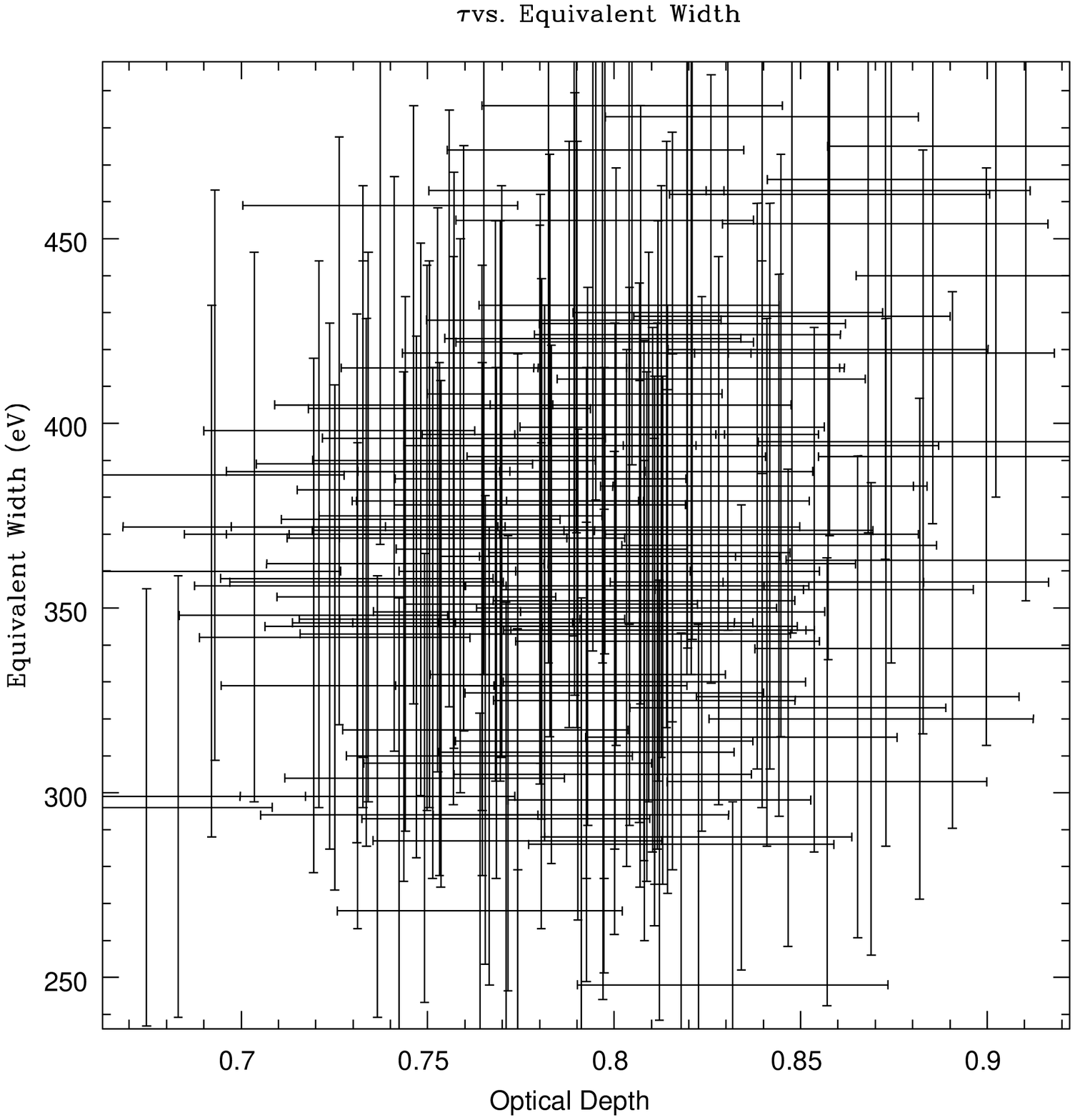,width=3.5in,angle=0}}
\caption{(a) The correlation between optical depth and the line
normalisation. (b) The lack of a correlation between optical depth and
the line equivalent width.  The figures are plotted from observation
10512-01-09-01, the longest soft state observation.  The plotted
errors show the 90\% confidence intervals.}
\end{figure*}

\begin{figure*}
\centerline{\epsfig{figure=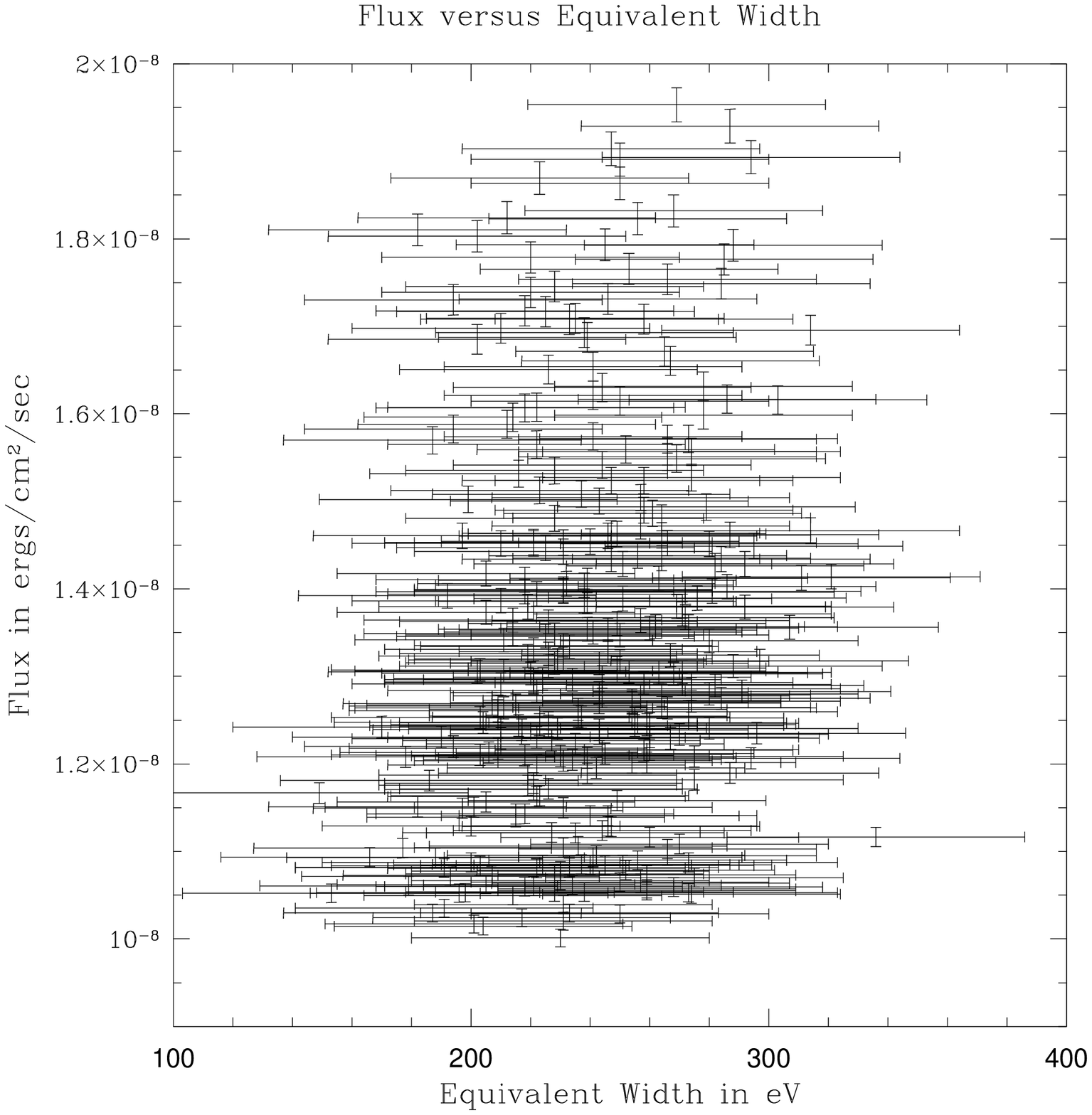,width=13cm,angle=0}}
\caption{The plot of flux versus equivalent width of the iron line for
observation 10412-01-01-00.  It is clear that the variations between
the two parameters have no correlation and that the variations in the
iron line equivalent width are consistent with the random errors.  The
plotted errors show the 90\% confidence intervals.}
\end{figure*}

\section{Comparison to the Hard State}

It has been previously shown that in the hard state of Cygnus X-1, the
variability must be driven by changes in the corona's properties
rather than changes in the properties of the seed photon distribution
(Maccarone, Coppi, \& Poutanen 2000; B\"ottcher 2001).  One of the key
pieces of evidence for this in hard state observations is that the
typical shot duration decreases with increasing energy.  This
observation rules out light-travel time models for the time lags seen
in the hard states of X-ray binaries (see e.g. Kazanas, Hua, \&
Titarchuk 1997).  The time lag spectra in the hard and soft state have
been shown to be similar to one another (Pottschmidt et al. 2000).
Additionally, we have now shown that the seed photon component is much
less variable that the coronal component of the spectrum, in
concordance with the results of Gilfanov et al. (2000) who found a
blackbody-type spectrum in their phase resolved spectroscopy that
showed up only at zero frequency.  Finally we present in Figure 5.8
the autocorrelations and cross-correlations of the three spectral
states and show that qualitatively, the same trends occur in all
three.  In all cases, the typical variability timescale is longer at
lower energies.  Since the variability is driven by changes in the
corona in all cases and the qualitative trend of typical variability
timescale versus energy is also the same, it seems likely that a
single origin might explain the variability in all three states.

\begin{figure*}
\centerline{\epsfig{figure=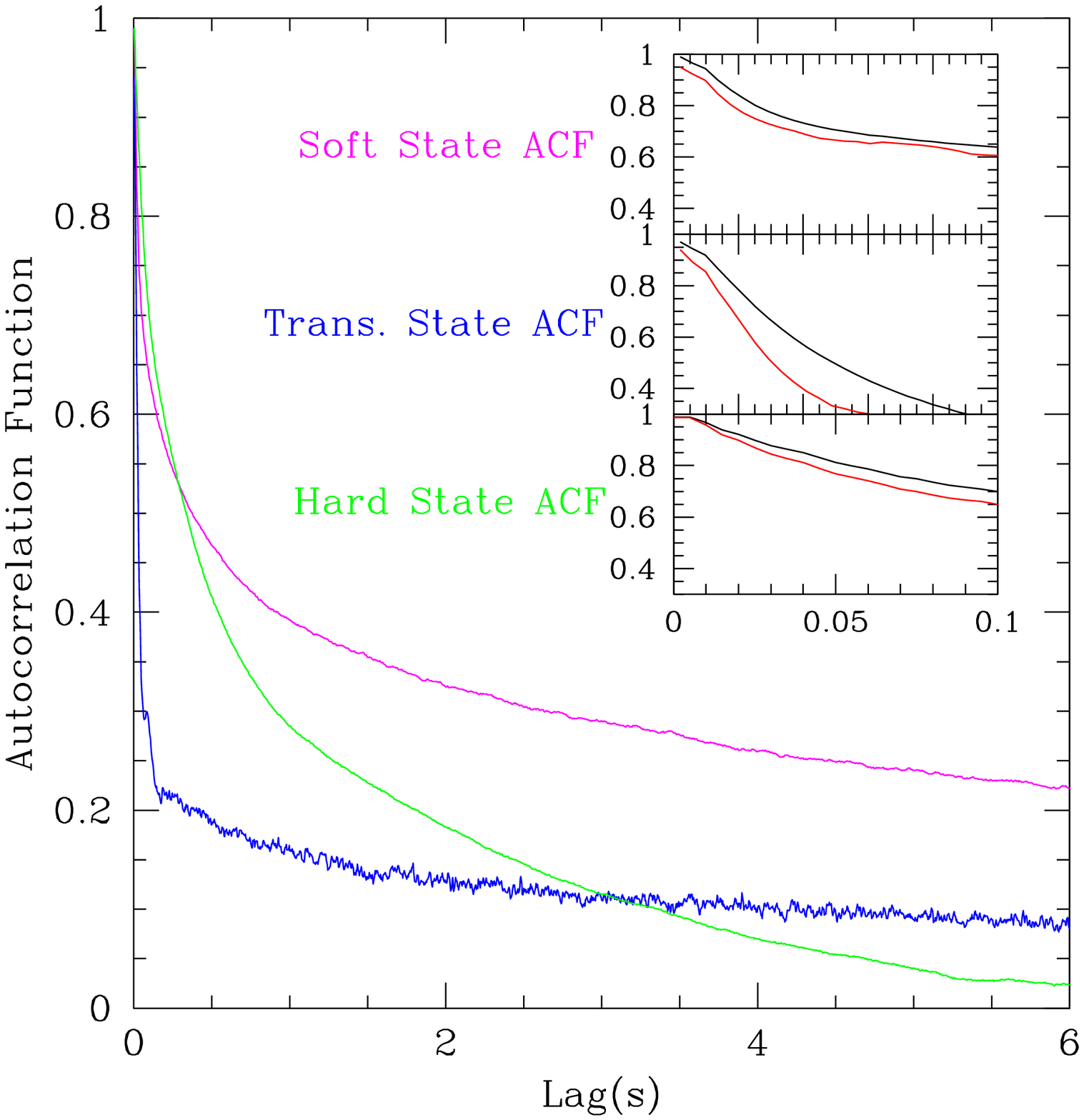,width=13 cm,angle=0}}
\caption{The autocorrelations in the three spectral states.  For the large graph, the soft state is in purple, the transition state in blue and the hard state in green.  For the inset boxes, the 2-5 keV ACF is in black while the 5-13 keV ACF is in red.}
\end{figure*}

\section{Summary} 

We find that the equivalent width of the iron line in Cygnus X-1
remains constant on 16 second timescales in the soft state, despite
changes in continuum flux by factors of order 3.  A constant
equivalent width indicates that the line intensity and the continuum
intensity vary together, i.e. the phase of the line variations is the
same as that of the continuum variations.  This observation suggests
that (1) the line flux level has now been definitively shown to
correlate with the continuum flux level on short timescales, (2) at
least 80\% of both the line and the continuum fluxes comes from within
about 16 light seconds of the black hole (most likely nearly 100\% of
the flux comes from these distances, but the fit errors preclude a
stronger statement) and (3) it is unlikely that major changes in the
accretion geometry occur during the soft state.

We find essentially the same results for the transition state.  There
is suggestive evidence that the iron line equivalent width evolves
over the transition state.  (In fact this almost certainly must be the
case since the equivalent widths in the soft state are slightly larger
than those in the transition state).  Thus we find again, that the
line flux is correlated with the continuum flux level, that the bulk
of the X-ray luminosity in both the line and in the continuum must
comes from within 16 light-seconds, and that there are unlikely to be
major geometry changes in the accretion flow in the transition state
on short timescales (i.e. less than 10000 seconds).

It is worth noting that the requirement that essentially all the iron
line flux comes from within 16 light seconds requires that the
emission comes from reflection off the accretion disc, and not off the
companion star or the outer portion of the stellar wind that makes the
accretion stream, since for the 5.6 day orbit and roughly 30
$M_{\odot}$ total mass of the Cyg X-1/HD 226868 system, the binary
separation between the black hole and its companion star is roughly
0.2 AU, or 100 light-seconds.  Since this is substantially larger than
the 16 light-seconds on which we observe the changes in the line and
continuum intensities, one would expect to observe an anti-correlation
between equivalent width and continuum intensity as well as a time
delay between continuum rises and line intensity rises if a
substantial portion of the line emission was due to reflection off the
companion star.  This finding is in accord with the observation from
ASCA that, in the hard state, the iron line equivalent width can drop
to less than 30 eV in the hard state (e.g. Ebisawa et al. 1996), which
provides an upper limit for the reprocessing effects of the companion
star and its wind.

Our two key results, (1) that no variability is detected in the seed
photon component on timescales where the hard component varies
strongly and (2) that the iron line flux variations track the
continuum flux variations, supply strong constraints on the nature of
the processes driving the X-ray variability.  One might expect to see
changes in the properties of the blackbody as a result of the energy
deposited in the disc by the Compton reflection process. However, the
corona's contribution to the total energetics of the system in the
soft and transition states is typically of order 20\%, so the soft
component dominates the overall energetics of the system (Frontera et
al.  2001).  Moreover, since the short timescale variability of the
coronal luminosity is only about 10\%, the seed photon component would
be expected to vary by only a few percent even if Compton reflection
deposited 100\% of the hard photons' energy into the disc.  Thus,
assuming the disk is intrinsically steady, one should not be too
surprised to see no evidence for the disc varying, assuming that the
disc is intrinsically steady. Both the steadiness of the disc
component and the correlation between $y$ and the line strength
suggest that the variability is instead driven by changes in the
properties of the corona, e.g., as in the spectral evolution model of
Poutanen \& Fabian 1999, rather than by changes in the soft photon
flux with a steady corona present (see e.g. Kazanas, Hua \& Titarchuk
1997; B\"ottcher \& Liang 1999).  The models of spectral variability
in accreting black holes have been developed almost entirely with the
hard state in mind, but the observations of Cygnus X-1 in the soft and
transition states are qualitatively consistent with what is seen in
the hard state of Cygnus X-1 in that coronal variations rather than
modulation of the soft component must be invoked (see e.g. Maccarone,
Coppi, \& Poutanen 2000).  Furthermore, our finding that the disc
appears intrinsically steady is in good agreement with observations of
other systems where the soft state spectra show essentially no power
law tails and the soft state lightcurves show essentially no
variability.

\section{Acknowledgements}

We wish to thank Mike Nowak and Raj Jain for useful discussions.  We
thank Charles Bailyn and Cole Miller for critical readings of this
manuscript and hope the readers of this work are grateful for their
insistence that the plots be made such that the data and model could
be distinguished more easily.  We are indebted to the anonymous
referee for numerous suggestions that improved the quality of this
work substantially.  This work has made use of data obtained from the
HEASARC online archive operated by NASA's Goddard Space Flight Center.

\label{lastpage}

\begin{thebibliography}{}
\bibitem{b1}Antonucci, R.R.J. \& Miller, J.S., 1985, ApJ, 297, 621
\bibitem{b2}Arnaud,  K.A.,  1996,  in  Astronomical Data  Analysis  Software  and Systems V,  ASP, Conf.   Ser., Vol.  101,  Jacoby, G., \&  Barnes, J.,
eds., (ASP: San Francisco), p. 17
\bibitem{b3}Basko, M., Sunyaev, R., \& Titarchuk, L., 1974, A\&A, 31, 249
\bibitem{b4}B\"ottcher, M., 2001, ApJ, 550, 963
\bibitem{b5}B\"ottcher, M. \& Liang, E.P., 1999, ApJL, 511, L37
\bibitem{b6}Bradt, H.V., Rothschild, R.E., \& Swank, J.H., 1993, A\&AS, 97, 355
\bibitem{b7}Ebisawa, K., Ueda, Y., Inoue, H., Tanaka, Y., \& White, N.E., 1996,ApJ, 467, 419
\bibitem{b8}Frontera, F. et al.  2001, ApJ, 546, 1027
\bibitem{b9}George, I.M., \& Fabian, A.C., 1991, MNRAS, 249, 352
\bibitem{b10}Gierlinski, M.,  Zdziarski, A.A., Poutanen, J.,  Coppi, P.S., Ebisawa,
K., \& Johnson, W.N., 1999, MNRAS, 309, 496
\bibitem{b11}Gilfanov, M., Churazov, E., \& Revnivtsev, M., 2000, MNRAS, 316, 923
\bibitem{b12}Kazanas, D., Hua, X.-M., \& Titarchuk, L., 1997, ApJ, 480, 735
\bibitem{b13}Lee, J.C., Fabian, A.C., Reynolds, C.S., Brandt, W.N., \& Iwasawa, K., 2000, MNRAS, 318, 857
\bibitem{b14}Maccarone, T.J., Coppi, P.S., \& Poutanen, J., 2000, ApJL, 537, L107
\bibitem{b15}Magdziarz, P. \& Zdziarski, A.A., 1995, MNRAS, 273, 837
\bibitem{b16}Mitsuda, K., Inoue, H., Koyama, K., Makishima, K., Matsuoka, M.,
Ogawara, Y., Suzuki, K., Tanaka, Y., Shibazaki, N., Hirano, T., 1984,
PASJ, 36, 741
\bibitem{b17}Nandra, K., George, I.M., Mushotzky, R.F., Turner, T.J., \& Yaqoob,
T., 1999, ApJL, 523, L17
\bibitem{b18}Nishimura, J., Mitsuda, K., \& Itoh, M., 1986, PASJ, 38, 819
\bibitem{b19}Nowak, M.A., 1995, PASP, 107, 1207
\bibitem{b20}Pottschmidt, K., Wilms, J., Nowak, M.A., Heindl, W.A., Smith, D.M., \&Staubert, R., 2000, A\&A, 357, 17L
\bibitem{b21}Poutanen, J., \& Fabian, A.C., 1999, MNRAS, 306, L31
\bibitem{b22}Shakura, N., \& Sunyaev, R., 1973, A\&A, 24, 337
\bibitem{b23}Wang, J.X., Zhou, Y.Y., Xu, H.G., \& Wang, T.G., 1999, ApJ, 516, L65
\bibitem{b24}Weaver, K.A. \& Reynolds, C., 1998, ApJL, 503, L39
\bibitem{b25}Yaqoob, T., Serlemitsos, P.J., Turner, T.J., George, I.M., \& Nandra,K., ApJL, 470, L27

\end{thebibliography}
\end{document}